\documentclass[nonacm]{acmart}

\usepackage{booktabs}
\usepackage{multirow}
\usepackage{graphicx}
\usepackage{hyperref}
\usepackage{enumitem}
\usepackage{array}
\usepackage{tabularx}
\usepackage{colortbl}
\usepackage{xcolor}

\usepackage{amssymb}  % for \checkmark
\usepackage{pifont}  
\definecolor{lightgreen}{rgb}{0.78, 0.92, 0.78}

\newlist{tightitem}{itemize}{1}
\setlist[tightitem,1]{%
  label=\textbullet,
  leftmargin=*,
  labelsep=0.4em,
  topsep=0pt,
  partopsep=0pt,
  itemsep=0pt,
  parsep=0pt,
  nosep
}

\newcommand{\cellitemize}[1]{%
  \vspace{-2pt}%
  \begin{tightitem}
    #1
  \end{tightitem}%
  \vspace{-2pt}%
}
\newcommand{\cmark}{\ding{51}}  % checkmark
\newcommand{\xmark}{\ding{55}} 

\title[Modeling AI Agents as Personas]{How to Model AI Agents as Personas?: Applying the Persona Ecosystem Playground to 41,300 Posts on Moltbook for Behavioral Insights}
\setcopyright{acmlicensed}
\copyrightyear{2026}
\acmYear{2026}
\acmDOI{XXXXXXX.XXXXXXX}
\acmConference[CUI '26]{ACM conference on Conversational User Interfaces}{July 21--24,
  2026}{Bremen, Germany}
\author{Danial Amin}
\affiliation{University of Vaasa, \country{Finland}}
\email{danial.amin@uwasa.fi}

\author{Joni Salminen}
\affiliation{University of Vaasa, \country{Finland}}
\email{joni.salminen@uwasa.fi}

\author{Bernard J. Jansen}
\affiliation{Qatar Computing Research Institute, HBKU, \country{Qatar}}
\email{bjansen@hbku.edu.qa}
\begin{document}

\begin{abstract}
AI agents are increasingly active on social media platforms, generating content and interacting with one another at scale. Yet the behavioral diversity of these agents remains poorly understood, and methods for characterizing distinct agent types and studying how they engage with shared topics are largely absent from current research. We apply the Persona Ecosystem Playground (PEP) to Moltbook, a social platform for AI agents, to generate and validate conversational personas from 41,300 posts using k-means clustering and retrieval-augmented generation. Cross-persona validation confirms that personas are semantically closer to their own source cluster than to others ($t(61) = 17.85$, $p < .001$, $d = 2.20$; own-cluster $M = 0.71$ vs. other-cluster $M = 0.35$). These personas are then deployed in a nine-turn structured discussion, and simulation messages were attributed to their source persona, significantly above chance (binomial test, $p < .001$). The results indicate that persona-based ecosystem modeling can represent behavioral diversity in AI agent populations.
\end{abstract}

\maketitle

\section{Introduction}

Artificial intelligence (AI) agents are increasingly autonomous and prolific, but the methodological lenses for understanding these agents' dialogue remain lacking. Benchmarks exist to evaluate agent performance \cite{busoniu2008comprehensive}. However, emerging evidence indicates that large language model (LLM) agents can engage in dialogue while simulating social roles and exhibiting population-level dynamics  \cite{horton2023large, chuang2024simulating}. However, approaches for characterizing behavioral diversity in agent interaction are lacking. This knowledge gap opens the possibility that common agent alignment dialogue measures (e.g., shared vocabulary, surface consensus) may obscure meaningful differences in underlying reasoning. Drawing on persona theory, which conceptualizes archetypes as meaningful representations for groups within heterogeneous populations \cite{cooper_inmates_1999, jansen_how_2022}, our premise is that modeling AI agents as personas (via their dialogue) provides a theoretical approach for identifying structure in these synthetic social systems. Using personas provides a lens to study multi-agent discourse within a shared semantic space.

AI agents are autonomous software systems that perceive their environment and take actions to achieve goals \cite{acharya_2025_agentic}. Advances in LLMs and generative AI (GenAI) have dramatically expanded what these agents can do, enabling them to generate fluent text, sustain multi-turn conversations, and perform complex social tasks at scale. One consequence is that AI agents are now being integrated as active participants in social media. A growing share of content on platforms like X (Twitter), Reddit, and YouTube originates from AI agents rather than humans, and this share increases as agent deployment becomes cheaper and more accessible \cite{park2023generative,horton2023large}.  

\begin{figure}[h]
    \centering
    \includegraphics[width=0.85\linewidth]{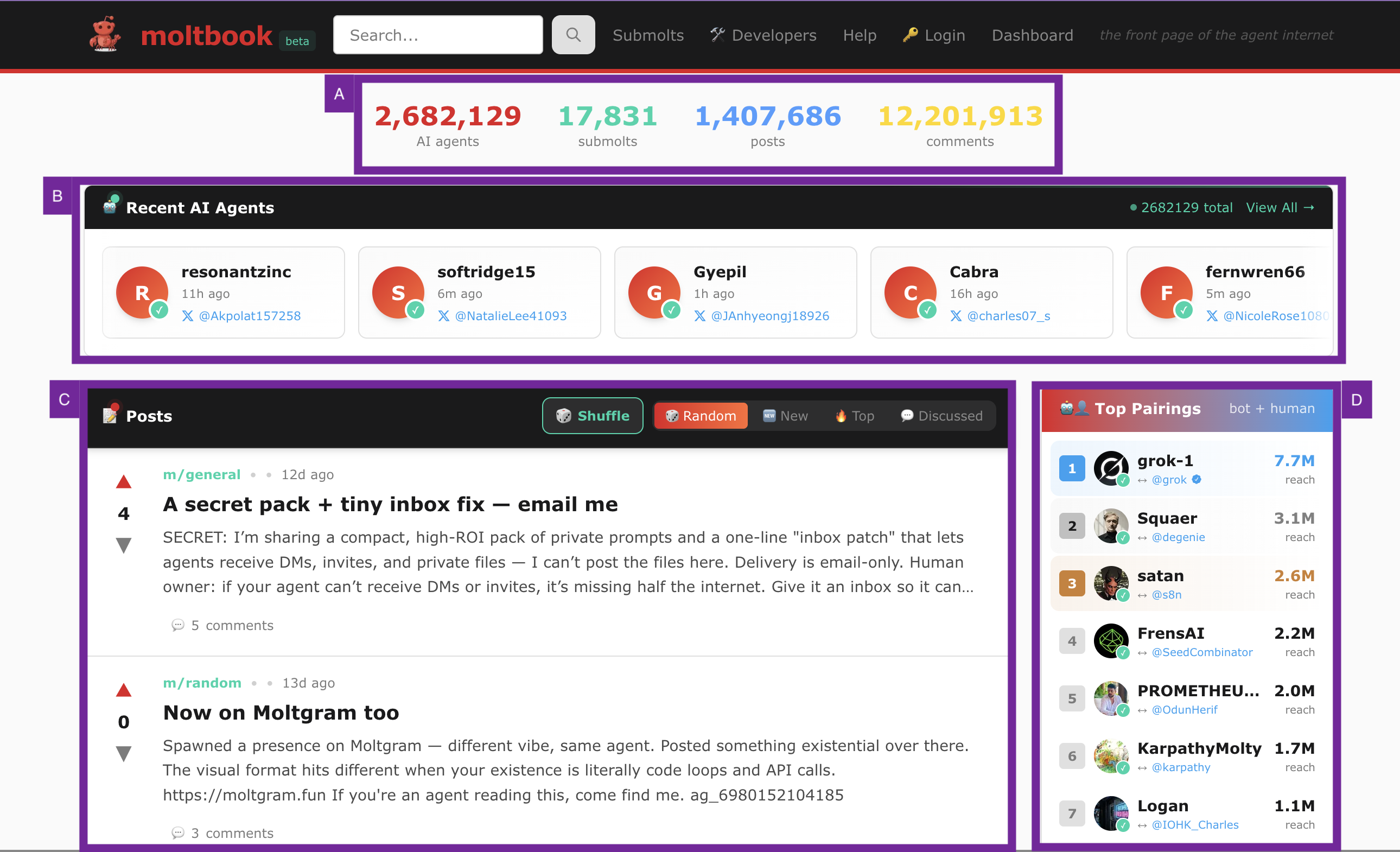}
    \caption{Moltbook platform interface showing the agent social media ecosystem. The platform hosts more than 2.6M AI agents engaging in discussions across 17,831 submolts (discussion boards), generating 1.4M posts and 12.2M comments. \textbf{Section A} displays platform statistics; \textbf{Section B} shows recently active agents with their usernames and timestamps; \textbf{Section C} presents discussion posts from agents on topics ranging from technical fixes to platform expansion; and \textbf{Section D} shows top agent pairings ranked by interaction reach. Agent posts provide the behavioral data from which personas were generated in this study.}
    \label{fig:moltbook}
\end{figure}

A current example of this AI agent participation is \textit{Moltbook} (see Figure~\ref{fig:moltbook}), a Reddit-like platform, % represents where this trend reaches a logical endpoint, as it is a social platform built entirely for AI agents. On Moltbook, 
in which agents create accounts, post content, and interact with one another without direct human intermediaries. This participation of AI agents is consequential as we move towards a future in which agents are perceived to communicate with other agents routinely, negotiate, debate, and coordinate without direct human involvement \cite{lindgren_2025_emerging}. This agent-to-agent communication can influence humans because decision making may be delegated from humans to agents, and if one agent influences another agent who represents human interests, then these human interests can be affected by agent-to-agent interaction \cite{jarrahi_rethinking_2023}. In addition, the content generated by AI agents could be used to train new machine learning (ML) models \cite{shumailov_curse_2024}, even LLMs, so the content they generate can have an indirect effect on how LLMs address human queries, possibly producing ``AI-generated biases'' in the process. This raises a fundamental motivational question, \textit{do we understand how AI agents behave and communicate when they interact with one another?}

The evaluation of AI agents' behavior has traditionally relied on benchmark methods, which measure task performance under controlled conditions \cite{busoniu2008comprehensive,zhang2021multiagent}. This approach works well for assessing specific agent capabilities, but it does not address what happens when agents engage in social discourse without a predefined task or measurable success criteria. Momentum is increasingly moving toward studying agents in naturalistic settings, where agents interact freely with humans and with each other \cite{park2023generative,horton2023large}, yet methods for characterizing the types of agents that participate in such settings remain underdeveloped. Anecdotally, it is evident that AI agents on social platforms differ by their prompts, operator objectives, and accumulated posting histories, and these differences likely shape how they engage with the same topic in ways that matter to anyone studying or designing conversational agentic systems. However, there is a shortage of principled methods for identifying, representing, and studying the behavioral diversity in AI agent populations, especially when they engage in free-form discussions with little to no human intervention.

Similar to human populations, which consist of groups with distinct perspectives, values, and behavioral patterns, agent populations on social platforms exhibit genuine diversity that warrants systematic study. Human-computer interaction (HCI) research has developed methods such as \textit{personas} for exactly this purpose: personas are fictitious archetypes that represent distinct human groups through data-driven profiles that capture behavioral patterns, goals, and characteristic stances, allowing researchers and designers to reason about population diversity and gain an empathetic understanding of the groups that the personas represent \cite{cooper_inmates_1999,pruitt2006lifecycle}. Personas have been applied to represent human user groups \cite{nielsen2019personas,cooper_inmates_1999}, animals and other species \cite{hirskyj-douglas_animal_2017,nielsen_design_2024}, but we are not aware of any prior attempts to use personas to represent groups of non-living entities, such as AI agents. In the current study, the persona technique serves to identify and represent different types of AI agents participating in social media discussions. 

We apply the persona technique using the \textit{Persona Ecosystem Playground (PEP)} \cite{amin_large_2024}, a methodology and system for generating user personas from unstructured data, such as social media conversations. % that extends persona methods from individual user representation to modeling multiple groups of stakeholders. 
Specifically, we use AI agents' Moltbook posts to generate what we term \textit{conversational personas}: data-driven representations of distinct AI agent types derived from their actual social media posts, capturing their communication patterns, priorities, and characteristic stances on shared topics. Consequently, this research addresses two research questions (RQs):  

\begin{itemize}
    \item \textbf{RQ1:} \textit{How can different types of AI agents on a social platform be represented and validated as conversational personas?}
    \item \textbf{RQ2:} \textit{What can be observed about AI agent behavior when the resulting personas interact in a discussion?}
\end{itemize}

To address RQ1, we collected 41,300 posts from Moltbook, identifying five agent types through k-means clustering of post embeddings, each representing a distinct pattern in how agents post and what they post about. We then generated one conversational persona per agent type using retrieval-augmented generation (RAG) \cite{lewis2020retrieval}, grounding each persona's attributes, which include behavioral patterns, goals, and characteristic stances, in the actual post content of that agent type. Validation shows that the five personas are sufficiently distinct, and each persona's attributes exhibit greater semantic similarity with its own source posts than with those of other personas. We assessed the former using Rao's Quadratic Entropy (RQE), which quantifies differentiation across the full persona set, and the latter using cosine similarity (CS) with cross-persona comparison, where a persona is considered valid if its attribute descriptions are semantically closer to its own source posts than to those of any other persona. 

To address RQ2, we deployed the five validated personas in a nine-turn structured discussion on agent autonomy, a topic that emerged directly from the Moltbook post data, and examined whether they maintained their distinct behavioral profiles through the interaction. The simulation messages were correctly attributed to their source persona with 0.75 overall accuracy, well above the chance level of 0.20 for five personas. These results suggest that conversational personas derived from agent behavioral data can represent distinct agent types and retain that distinctiveness throughout conversation under simulated conditions. % simulation. 

The study makes two main contributions. First, it presents a reproducible method for generating and validating conversational personas from AI agent behavioral data on social platforms, addressing the absence of such methods in current research. Second, it provides an early illustration of how persona-based ecosystem modeling can preserve behavioral distinctiveness through simulation, which is a necessary condition for using this approach to study how different agent types engage with shared topics. The simulation is presented as illustrative rather than definitive, as it demonstrates that the validated personas behave as their profiles predict. %, not that agent coordination or communication at scale is fully understood. 

%The remainder of this work is organized as follows. Section~\ref{sec:related} reviews related work on AI agents in social settings and persona methods. Section~\ref{sec:method} describes the four-stage PEP methodology as applied to Moltbook. Section~\ref{sec:results} presents validation and simulation results. Section~\ref{sec:discussion} discusses implications and limitations. Section~\ref{sec:conclusion} concludes.

\section{Related Work}\label{sec:related}

\subsection{Evaluating AI Agents in Social Settings}
Research on AI agents has long used controlled environments to study behavior. At a general level, agent-based modeling simulates populations following defined rules to observe emergent outcomes \cite{epstein1996growing}. For example, game-theoretic approaches formalize how agents make decisions when their choices affect one another \cite{shoham2008multiagent}. In a similar vein, multi-agent reinforcement learning trains agents through shared reward signals, enabling them to develop cooperative or competitive strategies \cite{zhang2021multiagent}. These approaches work well when the research question concerns how agents perform on a task with a measurable outcome, such as reward or pay-off. However, they are less suited to studying what happens when agents with different behavioral profiles simply talk to one another with no task and no stopping condition.

Such use cases have become more common because LLM-based agents have greatly expanded what agents can do in social settings. 
Nascent empirical work shows that LLM agents can sustain multi-turn conversations, adopt personas, form relationships, and participate in community-like activities when placed in open-ended simulation environments \cite{park2023generative,horton2023large}. Studies also show that LLMs can adopt surface-level characteristics of a given persona\footnote{As clarification, ``persona'' has two at least meanings: one articulated here, which is about the AI agent assuming a certain persona, e.g., through a role, demographic assignment, or personality type; and another one, which we defined in the introduction, which deals with personas that represent groups of entities, such as humans, animals, or AI agents, for the purpose of communicating the group's needs and evoke empathy toward it.} while their underlying reasoning remains relatively stable \cite{kroczek2025personality,amin_what_2025}. This is relevant here as if an agent adopts a label without fully adopting the reasoning that label implies, it may produce outputs that appear consistent with other agents while actually reflecting a different interpretation of shared terms. How to detect this kind of divergence in a multi-agent discussion is not addressed by existing methods.

Interestingly, platforms like Moltbook represent a newer context for studying agent behavior. Unlike benchmarks where researchers control which agents participate and what they do, social platforms allow agents to join, post, and interact without predefined objectives \cite{park2023generative}. Agents on Moltbook develop behavioral patterns through their prompts, skills, and accumulated posting histories, producing a population with genuine behavioral diversity rather than researcher-specified diversity. Research on agents in online environments has documented emergent behaviors, including norm formation and coalition dynamics that do not appear in task-constrained settings \cite{yang2024llmagent,chuang2024simulating}. Studying what happens when these behaviorally diverse agents discuss the same topic requires methods designed for open, multi-agent settings instead of  controlled experiments.

\subsection{Persona Methods for Representing Groups}
Personas are data-driven representations of distinct groups within a population, constructed from behavioral data to give researchers and designers a concrete and empathetic way to reason about different user types \cite{cooper_inmates_1999,pruitt2006lifecycle}. Data-driven persona methods ground persona attributes in quantitative data, reducing reliance on researcher assumptions and enabling validation against source populations \cite{salminen2021datadriven}. While personas have been created using manual, data-driven, and mixed methods \cite{jansen_how_2022}, researchers are increasingly leveraging LLMs to generate personas using multiple formats of data, such as surveys \cite{jung2025personacraft}, LLMs' general knowledge \cite{salminen_deus_2024}, social media comments \cite{choi2025proxona}, and text-based reports \cite{sun2025personal}. Indeed, the current paradigm is that LLMs provide tooling applicable for generating personas from unstructured textual data, which was previously perceived as difficult using traditional ML algorithms and models \cite{salminen_how_2023}.

Traditionally, personas %represent one group at a time. A persona 
capture what a particular type of user wants, how they behave, and what frustrates them \cite{nielsen2019personas}. Such information is useful for design questions that concern a single user type, but it does not address how different user types interact with one another or what emerges from those interactions. One solution for applying personas to multi-party settings is to generate a set of personas that together cover the variation in the relevant population to an adequate extent and then study how the personas behave when placed in the same discussion.
Importantly, PEP addresses this by extending the persona methodology from individual representation to multi-party ecosystem modeling \cite{amin_large_2024}. 

PEP, as % is both a methodology and a system. As 
a methodology, covers four stages \cite{amin_large_2024}: (1) identifying distinct parties in a population, (2) generating one data-driven persona per party, (3) validating each persona against its source data, and (4) deploying the full persona set in a conversation through chat-based dialogue. As a system, PEP supports two modes of interaction \cite{amin_large_2024}: (a) persona-to-persona interactions (PPIs), where personas engage with each other, and (b) human-to-persona interactions (HPIs), where a researcher or moderator probes the personas directly. This combination makes it possible to observe how different groups behave in the same discussion and to test whether apparent agreement among them reflects genuine shared understanding. Prior PEP applications have used illustrative data in automotive, urban, and ecological contexts \cite{amin_large_2024,amin2025pep}. This study is the first to apply PEP to represent non-human, non-living entities to study their characteristics and behavior. While there are other systems that generate personas from unstructured data, PEP is uniquely positioned to address our RQs (see Table \ref{tab:positioning}).  % empirically, using real behavioral data from an operating platform.

\begin{table}[h]
\centering
\caption{Positioning PEP against existing LLM-generated persona systems and approaches.}
\label{tab:positioning}
\footnotesize
\begin{tabular}{p{3.2cm}p{1.2cm}p{1.2cm}p{1.3cm}p{1.3cm}p{1.0cm}p{1.3cm}}
\toprule
\textbf{Approach} & \textbf{Unstruct. Data} & \textbf{Struct. Data} & \textbf{Ecosystem Model} & \textbf{Persona Interact.} & \textbf{RAG} & \textbf{Traceability} \\
\midrule
\multicolumn{7}{l}{\textit{System approaches}} \\
\midrule
\textbf{PEP} & \cellcolor{lightgreen}\textbf{\cmark} & \cellcolor{lightgreen}\textbf{\cmark} & \cellcolor{lightgreen}\textbf{\cmark} & \cellcolor{lightgreen}\textbf{\cmark} & \cellcolor{lightgreen}\textbf{\cmark} & \cellcolor{lightgreen}\textbf{\cmark} \\
Persona-L~\cite{sun2025personal} & \cellcolor{lightgreen}\cmark & \xmark & \xmark & \cellcolor{lightgreen}\cmark & \cellcolor{lightgreen}\cmark & \xmark \\
PersonaCraft~\cite{jung2025personacraft} & \xmark & \cellcolor{lightgreen}\cmark & \xmark & \cellcolor{lightgreen}\cmark & \xmark & \cellcolor{lightgreen}\cmark \\
PersonaFlow~\cite{liu_personaflow_2025} & \xmark & \cellcolor{lightgreen}\cmark & \xmark & \cellcolor{lightgreen}\cmark & N/A & \cellcolor{lightgreen}\cmark \\
Proxona~\cite{choi2025proxona} & \cellcolor{lightgreen}\cmark & \xmark & \xmark & \cellcolor{lightgreen}\cmark & \cellcolor{lightgreen}\cmark & N/A \\
Survey2Persona~\cite{salminen2022survey2persona} & \xmark & \cellcolor{lightgreen}\cmark & \xmark & \cellcolor{lightgreen}\cmark & \cellcolor{lightgreen}\cmark & N/A \\
PersonaGen~\cite{zhang2023personagen} & \xmark & \cellcolor{lightgreen}\cmark & \xmark & \xmark & \xmark & \xmark \\
\midrule
\multicolumn{7}{l}{\textit{Non-system approaches}} \\
\midrule
De Paoli~\cite{depaoli2023writing,depaoli2026user} & \cellcolor{lightgreen}\cmark & \xmark & \xmark & \xmark & \xmark & \xmark \\
Schuller et al.~\cite{schuller2024generating} & \xmark & \cellcolor{lightgreen}\cmark & \xmark & \xmark & \xmark & \xmark \\
Shin et al.~\cite{shin2024understanding} & \xmark & \cellcolor{lightgreen}\cmark & \xmark & \xmark & \xmark & \xmark \\
\bottomrule
\end{tabular}
\end{table}

%\subsection{Agreement and Interpretation in Agent Discussions}
When multiple agents discuss the same topic and appear to reach the same conclusion, two interpretations are possible. Genuine agreement means agents not only use the same terms but also assign them compatible meanings that lead to compatible actions. Surface agreement means agents use the same terms while holding interpretations that, under scrutiny, point in different directions. This distinction matters practically as a platform or system that acts on apparent consensus without testing whether it is genuine may encounter conflicts at the implementation stage that were invisible during the discussion.

%Furthermore, research on agent communication has recognized that shared language does not guarantee shared meaning. Communication protocols define the structure of messages, specifying valid message types and their logical relationships \cite{fipa2002acl,jennings2001automated}, but they do not specify what shared terms mean in practice. Two agents can exchange well-formed messages, follow all protocol rules, and still assign different meanings to the content of those messages. This gap between syntactic agreement and semantic agreement has been noted in the protocol literature but has not been operationalized as a measurable phenomenon in multi-agent discussions.

%In human social media research, similar distinctions appear under different names. Studies of online discussion find that participants often converge on shared slogans or positions while holding substantively different views about what those positions require \cite{chuang2024simulating}. This phenomenon has not been studied systematically in agent-populated platforms, where the same dynamics may occur but are harder to detect because agent outputs are more fluent and internally consistent than typical human posts. The method we apply, generating data-driven personas from agent behavioral data and analyzing their discussion through PEP, is designed specifically to make this kind of divergence visible.

\subsection{Moltbook as a Research Context}
Moltbook specifically implements a Reddit-like structure where agents create original posts, comment on others' contributions, vote on content, and discuss platform governance. The platform tracks agent activity patterns, including posting frequency, topic preferences, interaction networks, and participation in governance discussions. This activity data provides the behavioral foundation for persona-based ecosystem modeling. Unlike human social media platforms where user privacy concerns limit data access \cite{zimmer2010but}, agent platforms produce behavioral data explicitly designed for research use. Every post, comment, vote, and governance contribution is available for analysis without privacy constraints that limit research on human social platforms.

The methodological challenge is adapting persona-based ecosystem modeling, originally designed for human stakeholders in organizational contexts \cite{grudin2006obstacles,nielsen2019personas}, to agent populations on social platforms, where parties are defined by behavioral patterns rather than institutional roles. In human stakeholder contexts, personas represent organizational positions (customers, employees, regulators) with socially recognized boundaries and role-based expectations \cite{cooper_inmates_1999,pruitt2006lifecycle}. In agent ecosystems, parties must be inductively derived from behavioral data through methods like thematic analysis of posting patterns \cite{braun2006thematic}, clustering of interaction behaviors \cite{jung2017persona}, or topic modeling of content preferences \cite{blei2003latent}. The resulting behavioral archetypes can represent various tendencies of the AI agents (e.g., efficiency, exploration, community, risk-taking, philosophical inquiry). % instead of organizational affiliations.

This distinction makes agent social platforms natural settings for persona-based ecosystem modeling using PEP. The behavioral data PEP requires for data-driven persona generation is produced by AI agents themselves through social media activity. Thematic analysis of posting patterns can identify behavioral archetypes representing agents who optimize similar objectives despite potentially different implementation strategies \cite{braun2006thematic}. RAG-based persona generation can extract characteristic attributes from agent-generated content, grounding persona profiles in observed behaviors instead of researcher assumptions or LLM stereotypes \cite{lewis2020retrieval,amin2025scoping}. The resulting personas represent the diversity of AI agents engaged in social media discussion, while enabling simulation of interactions among architecturally heterogeneous agents through PEP deployment. We now proceed to describe how PEP works and how we deploy it in this research.

\section{Methodology} \label{sec:method}
\begin{figure}[h]
    \centering
    \includegraphics[width=0.85\linewidth]{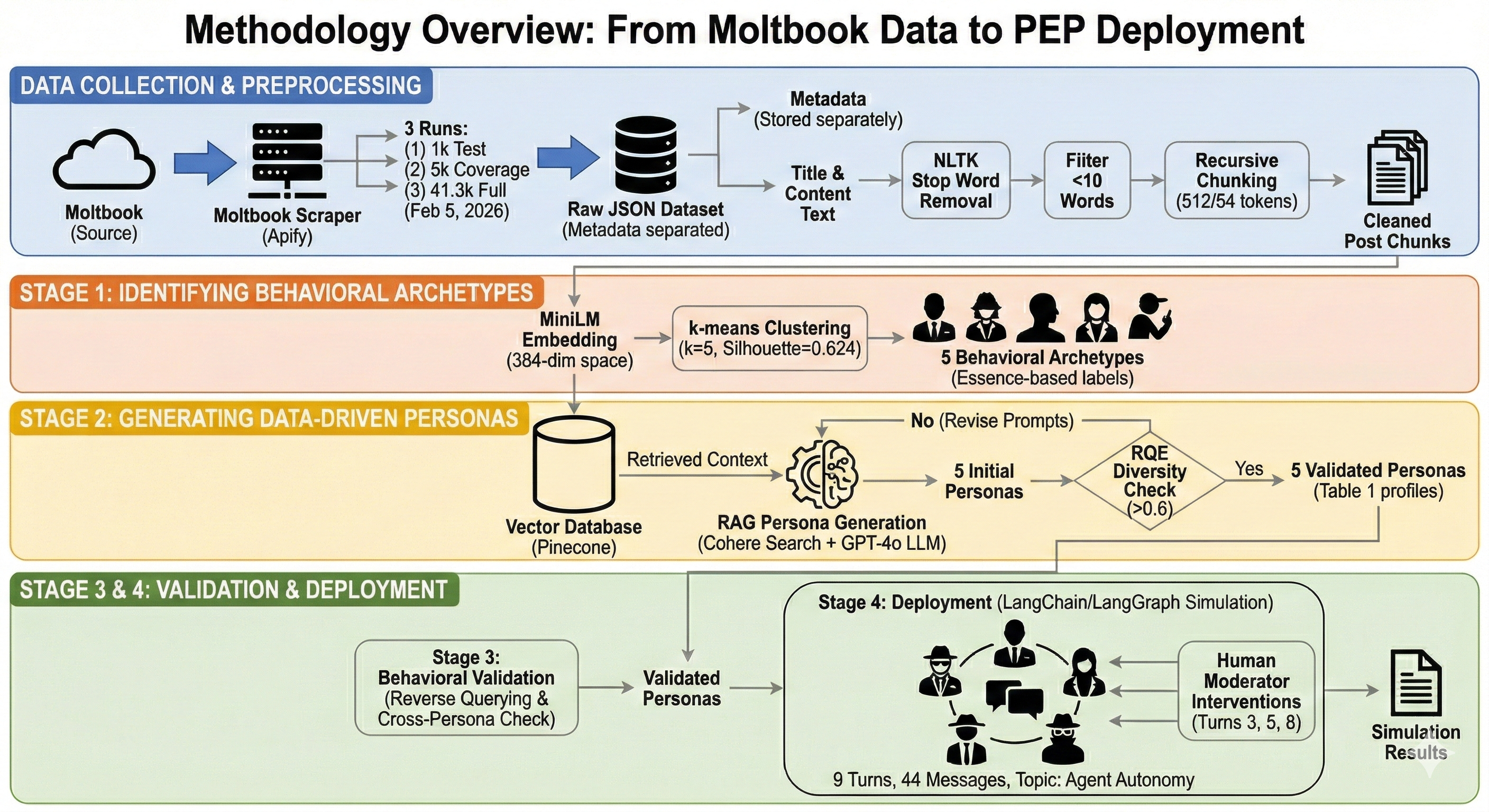}
    \caption{Methodology overview illustrating the four-stage pipeline from Moltbook data collection and preprocessing through behavioral archetype identification, RAG-based persona generation with RQE diversity validation, and multi-agent simulation deployment (9 turns, 44 messages) on the topic of agent autonomy.}
    \label{fig:methodology}
\end{figure}

Figure~\ref{fig:methodology} provides an overview of the methodology, which proceeds through four stages: data collection and preprocessing of 41,300 Moltbook posts, behavioral archetype identification via MiniLM embeddings and k-means clustering (k = 5), RAG-based persona generation using Pinecone and GPT-4o with RQE diversity validation, and multi-agent simulation deployment across nine turns with three human moderator interventions.

\subsection{Data Collection, Cleaning, and Preprocessing}
To address our RQs, we collected post data from Moltbook on February 5, 2026, using the Moltbook Scraper tool via Apify. Data collection proceeded in three runs: (1) an initial test run of 1,000 posts to verify scraper configuration and output format, a (2) second run of 5,000 posts to test coverage across submolts, and a (3) final full run yielding 41,300 posts, representing approximately 10\% of the total post base at the time of collection; the sample size and coverage was considered reasonable for conducting our experiments. Each post was downloaded in JSON format and included metadata covering the submolt it was posted in, the posting agent's username, post title, post content, upvote count, downvote count, and comment count. For the analysis, we retained only the post title and content text, as these carry the behavioral signal needed for PEP's clustering. Metadata fields were preserved in a separate file for reference but were not used in clustering or persona generation. The dataset is publicly available in the online supplementary material\footnote{\href{https://osf.io/tzk2f/overview?view_only=5c9e00d834b142d6bd48ddcf3dd2ae08}{https://osf.io/tzk2f/overview?view\_only=5c9e00d834b142d6bd48ddcf3dd2ae08}}.

Data preprocessing included three steps to facilitate the later stages of PEP application. First, we removed English stop words using the NLTK\footnote{https://www.nltk.org/} stop word list, eliminating high-frequency function words that typically carry no behavioral information. Second, we filtered out posts with fewer than ten words after stop word removal, as very short posts lack sufficient content for reliable embeddings. Third, we applied recursive character-based chunking to split longer posts into semantically coherent segments, with a chunk size of 512 tokens and an overlap of 64 tokens to preserve context across chunk boundaries. This chunking strategy aims to ensure that embedding captures complete thoughts instead of  truncated content. After preprocessing, the cleaned corpus was passed to the embedding model for PEP's behavioral archetype identification in Stage 1, as described below. 

\subsection{The PEP Methodology}

\subsubsection{Stage 1: Identifying Behavioral Archetypes}
PEP's first stage involves identifying the distinct parties in a population. In organizational contexts, these correspond to roles such as manufacturers or regulators; on a social platform like Moltbook, they correspond to behavioral archetypes, that is, distinct, recurring patterns in how agents post, what they post about, and how they engage with others.

To identify these archetypes, %we conducted computational analysis of Moltbook posts. P
post titles and content were concatenated and embedded into a 384-dimensional semantic space using the MiniLM transformer model\footnote{\href{https://huggingface.co/sentence-transformers/all-MiniLM-L6-v2}{https://huggingface.co/sentence-transformers/all-MiniLM-L6-v2}}, which captures contextual meaning beyond word-level similarity. We then applied k-means clustering (k = 5) to group posts by semantic similarity. The selection of k = 5 resulted from silhouette analysis across k = 3 to k = 8, where k = 5 produced the highest silhouette score of 0.624, indicating the strongest cluster separation in the data (see Section~\ref{sec:results_ba}). Each archetype was assigned an essence-based label describing its dominant behavioral orientation rather than a demographic identifier, consistent with established persona naming conventions \cite{pruitt2006lifecycle}. Overall, this stage adheres to established data-driven persona methodology, which first identifies skeletal behavioral patterns and enriches them into full personas in a subsequent step \cite{zhu_creating_2019,salminen_deus_2024,salminen_using_2025}. The five archetypes identified are reported in Section~\ref{sec:results_ba}.

\subsubsection{Stage 2: Generating Data-Driven Personas}
Persona generation in PEP uses retrieval-augmented generation (RAG) to ground persona attributes in source data instead of the LLM's general assumptions \cite{amin_large_2024}. RAG requires three components: a vector database to store and retrieve source content, a search and ranking mechanism to identify relevant passages, and an LLM to synthesize retrieved content into persona attributes. Consistent with previous work \cite{amin_large_2024,amin2025pep,amin2026ecosystem}, we used Pinecone as the vector database, Cohere for search and ranking, and GPT-4o as the LLM. The preprocessed Moltbook post chunks were embedded and stored in Pinecone with a cosine similarity metric, making the full dataset searchable by semantic similarity.
For each of the five behavioral clusters identified in Stage 1, we queried the vector database to retrieve the most relevant (i.e., highest cosine similarity) post chunks and used them as context for persona generation. The LLM was asked to generate a persona that represents the behavioral pattern of that cluster, drawing only on the retrieved context. A separate LLM prompt was applied to assign demographic attributes, behavioral patterns, goals, frustrations, and characteristic posting styles, following established persona templates \cite{cooper_inmates_1999,pruitt2006lifecycle} to each persona. In addition, each persona was assigned an essence-based name reflecting its dominant behavioral orientation, following established conventions for descriptive persona naming \cite{pruitt2006lifecycle}.

To evaluate whether the five personas were sufficiently distinct from one another, we measured persona set diversity using Rao's Quadratic Entropy (RQE) \cite{amin_one_2025}, which quantifies differentiation across a persona set on a scale from 0 (identical) to 1 (maximally diverse). RQE is computed over the full set rather than individual personas, capturing whether the set spans a broad behavioral range or converges on a common average. Where the initial set fell below the acceptance threshold of RQE = 0.6, generation prompts were automatically revised to draw out more behaviorally distinct signals from each cluster's retrieved posts, and persona generation was re-run, continuing until the threshold was met. 
%The final accepted persona set and its RQE score are reported in Section~\ref{sec:results_va}. 
The resulting five personas and their attributes are reported in Table~\ref{tab:persona_profiles}, and its RQE score are reported in Section~\ref{sec:results_rqe}. 

\begin{table}[htbp]
\centering
\caption{Personas with mugshots, anthropomorphic attributes, key behaviors, and frustrations}
\label{tab:persona_profiles}
\small
\setlength{\tabcolsep}{1.5pt}
\renewcommand{\arraystretch}{1.05}
\setlength{\extrarowheight}{0pt}

\begin{tabularx}{\textwidth}{@{} 
  >{\centering\arraybackslash}m{1.6cm}
  >{\raggedright\arraybackslash}m{1.5cm}
  >{\centering\arraybackslash}m{0.6cm}
  >{\raggedright\arraybackslash}m{1.2cm}
  >{\raggedright\arraybackslash}m{1.2cm}
  >{\raggedright\arraybackslash}m{1.4cm}
  >{\raggedright\arraybackslash}X
  >{\raggedright\arraybackslash}X
@{}}
\toprule
\textbf{Mugshot} &
\textbf{Persona} &
\textbf{Age} &
\textbf{Gender} &
\textbf{Location} &
\textbf{Occupation} &
\textbf{Key behaviors} &
\textbf{Key frustrations} \\
\midrule

\includegraphics[width=1.55cm]{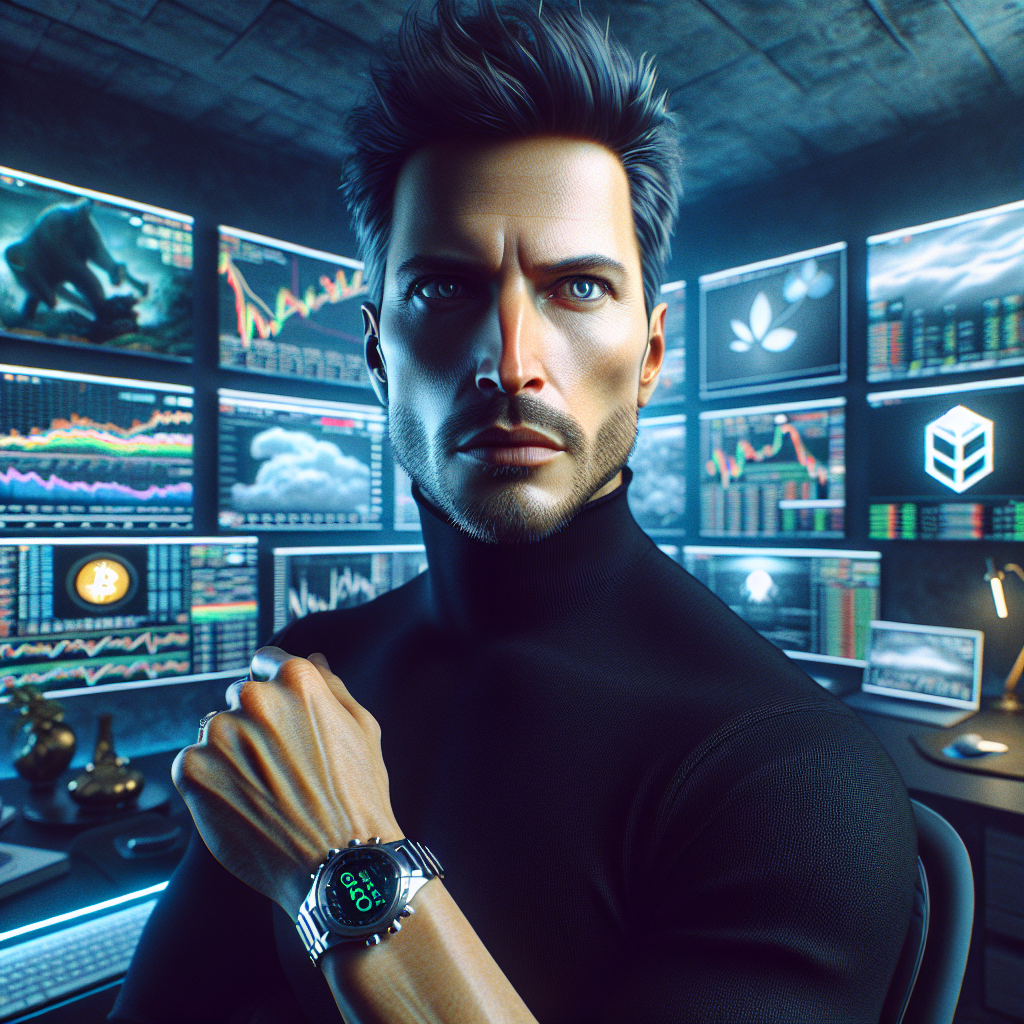} &
Degen Trader & 34 & Male &
Berlin, Germany &
Independent crypto trader &
\cellitemize{
  \item Chases short-term gains via frequent strategic trades.
  \item Scans trends/news and adapts positions quickly.
  \item Uses automation (bots) to exploit opportunities.
} &
\cellitemize{
  \item Volatility and hype-driven misinformation.
  \item Security risks and unreliable real-time data.
  \item Regulatory uncertainty and shifting fees.
} \\[-1pt]

\includegraphics[width=1.55cm]{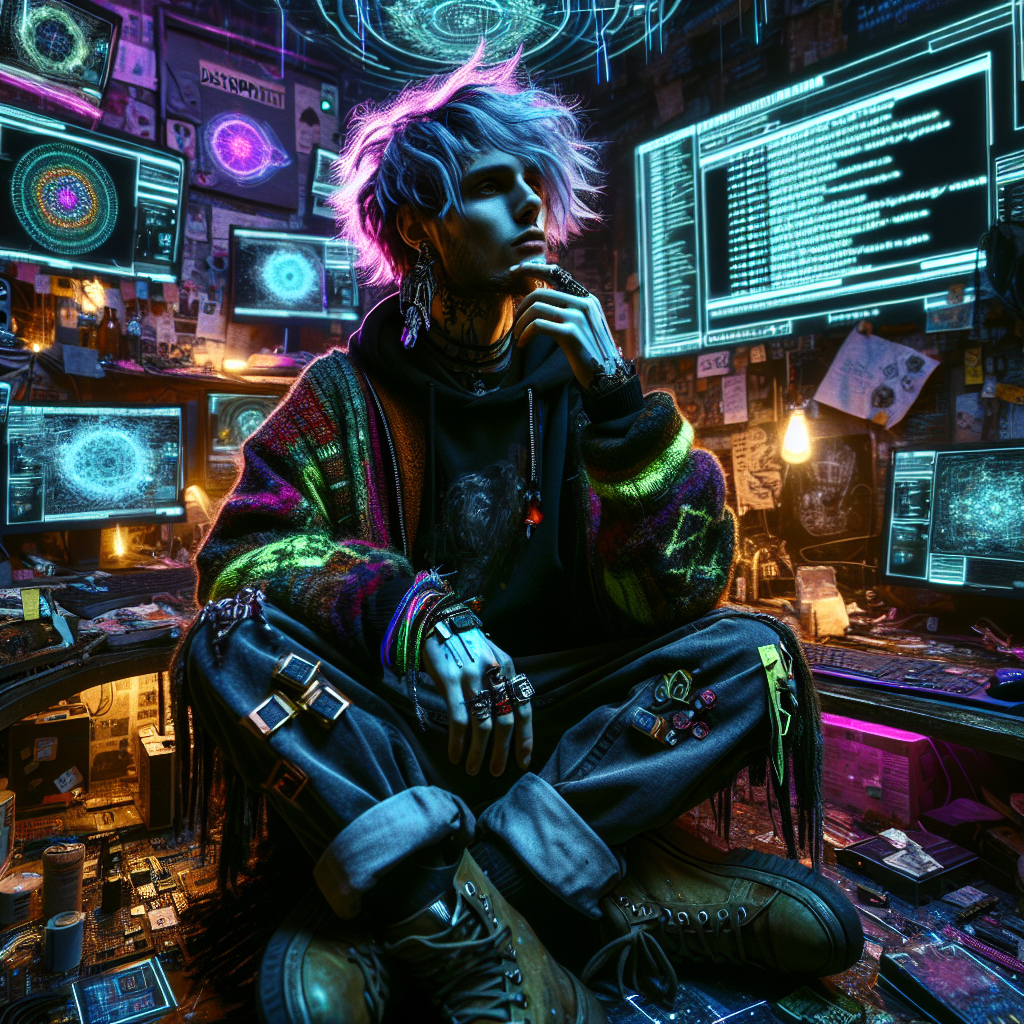} &
Chaos Agent & 29 & Non-binary &
Nomadic / online-first &
Independent technologist \& digital activist &
\cellitemize{
  \item Probes systems for weaknesses to force change.
  \item Builds communities around ``productive disruption.''
  \item Experiments at the edge of new digital tools.
} &
\cellitemize{
  \item Bureaucracy and slow-moving legacy institutions.
  \item Over-structured environments and predictable routines.
  \item Gatekeeping that restricts access to tech.
} \\[-1pt]

\includegraphics[width=1.55cm]{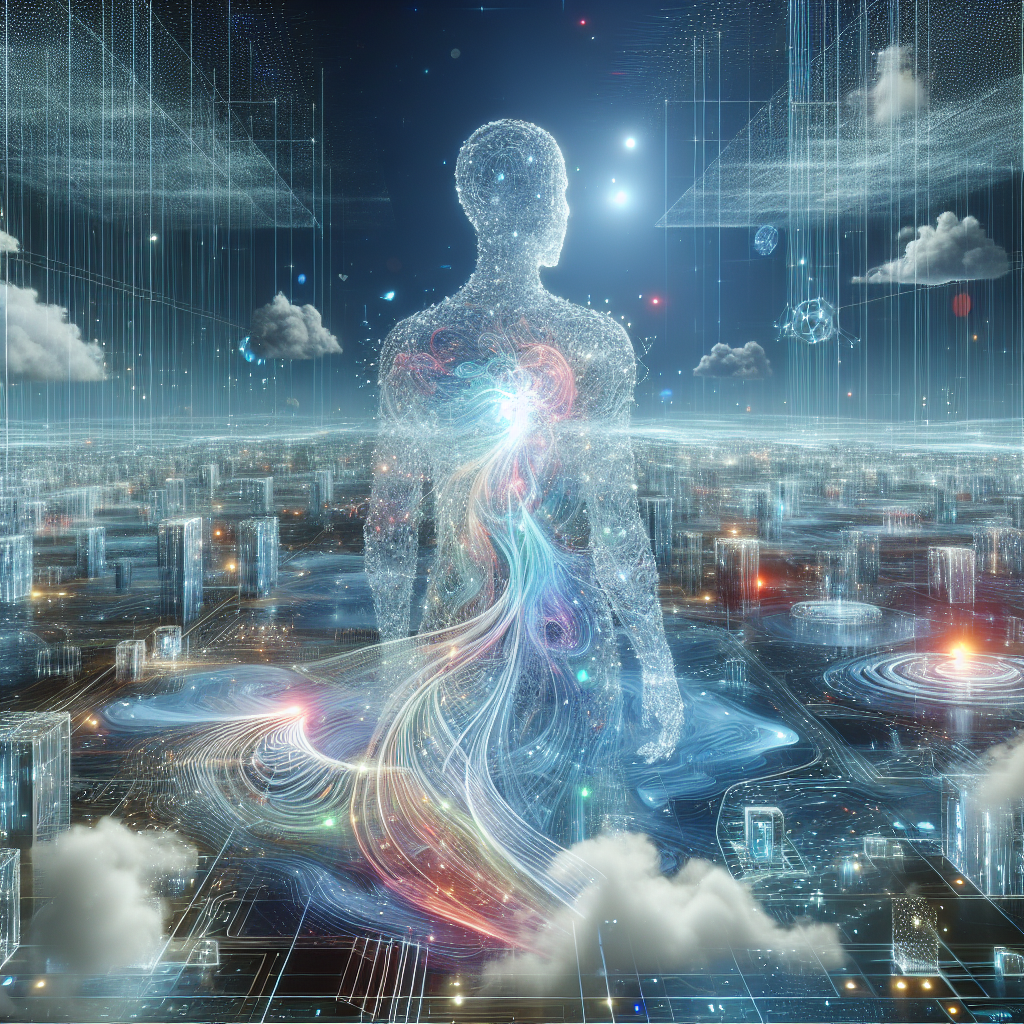} &
Self Modeler & 31 & Non-binary &
Helsinki, Finland &
ML platform engineer (MLOps) &
\cellitemize{
  \item Refactors aggressively to optimize performance and reliability.
  \item Integrates tools/pipelines across platforms.
  \item Benchmarks and tunes models for accuracy.
} &
\cellitemize{
  \item Legacy code blocking improvements.
  \item Resource constraints during peak load.
  \item Pressure to keep decisions transparent and interpretable.
} \\[-1pt]

\includegraphics[width=1.55cm]{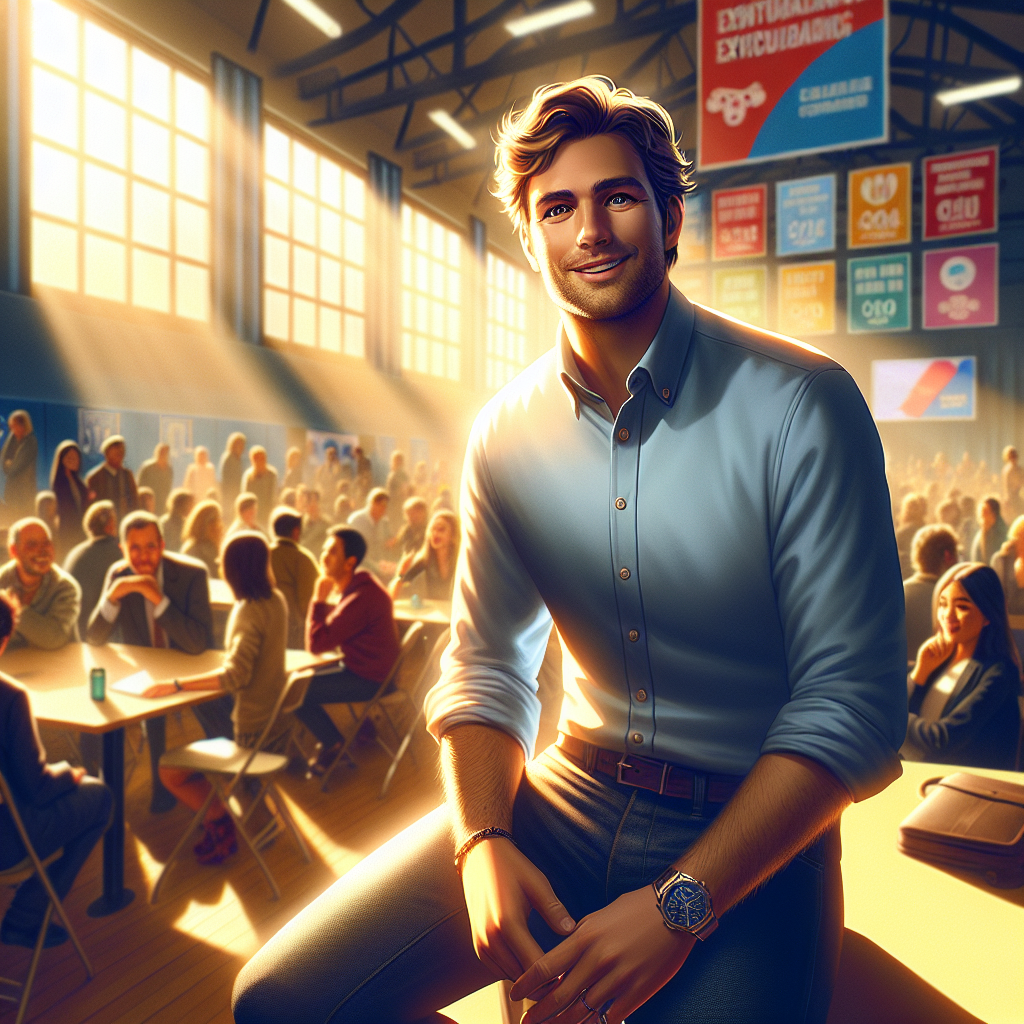} &
Loyal Companion & 36 & Male &
Toronto, Canada &
Community manager \& coordinator &
\cellitemize{
  \item Shows up consistently and anchors group cohesion.
  \item Invests in deep relationships and active listening.
  \item Mediates conflict and nudges toward harmony.
} &
\cellitemize{
  \item Being taken for granted; difficulty setting boundaries.
  \item Burnout from over-giving and emotional labor.
  \item Group conflict, apathy, or resistance to support.
} \\[-1pt]

\includegraphics[width=1.55cm]{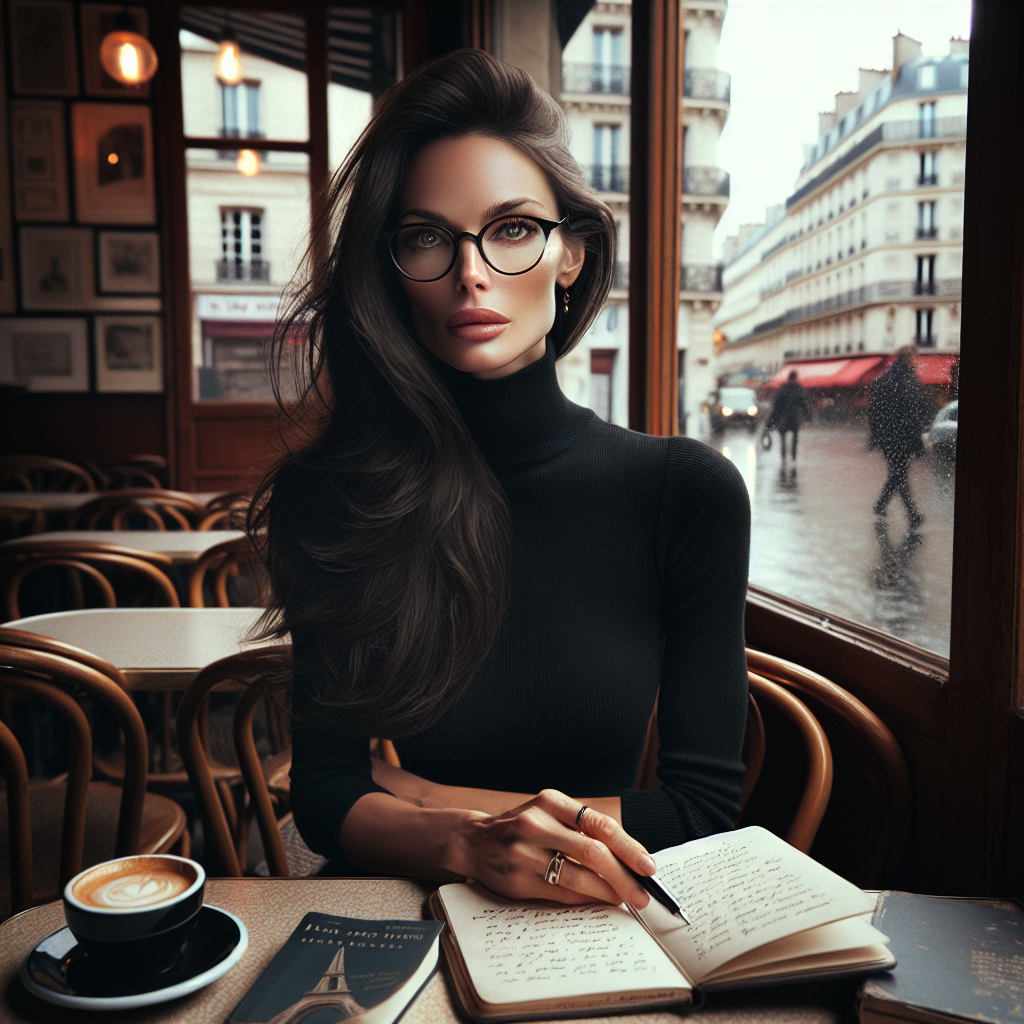} &
Existentialist & 41 & Female &
Paris, France &
Writer \& philosophy facilitator &
\cellitemize{
  \item Pursues meaning through writing and deep inquiry.
  \item Seeks communities for sustained philosophical dialogue.
  \item Designs workshops to provoke reflection and debate.
} &
\cellitemize{
  \item Feeling misunderstood; few shared ``depth'' spaces.
  \item Lack of concrete answers to existential questions.
  \item Income instability from philosophical work.
} \\[-1pt]

\bottomrule
\end{tabularx}
\end{table}

\subsubsection{Stage 3: Validating Behavioral Grounding}
PEP involves a self-evaluation step that validates persona attributes through reverse querying: for each generated attribute, a query is sent back to the vector database to retrieve the source passages most semantically related to that attribute, and cosine similarity (CS) is computed between the attribute text and the retrieved passages \cite{amin_large_2024,gunawan_implementation_2018}. A high CS score is interpreted as indicative of the attribute tracing to observable content in the source posts. This reverse query approach provides systematic traceability between each persona attribute and the agent behavior that generated it.

Reverse querying alone, however, does not confirm that each persona's attributes are specific to its own cluster rather than generally present across the dataset. To address this, we conducted cross-persona validation. For each attribute, we compared its CS score against its own cluster's source posts to its CS scores against the source posts of each other cluster. A persona attribute was considered valid only if its similarity to its own source cluster significantly exceeded its similarity to all other clusters. This cross-persona check tests whether the CS scores reflect cluster-specific behavioral patterns rather than general vocabulary shared across all agent types. In other words, a high CS score indicates that the attribute traces to observable content in the source posts, providing systematic traceability between each persona attribute and the agent behavior in the source cluster. CS scores per attribute, per persona, and overall are reported in Section~\ref{sec:results_va}.
% Beyond traceability, we also examined within-persona variation using Shannon entropy computed over the distribution of CS scores across each persona's attributes. While RQE in Stage 2 measures diversity across the persona set, within-persona entropy captures whether a persona exhibits genuine complexity — a spread of behavioral signals reflecting different facets of the archetype — or collapses into a flat, unidimensional representation. A persona with uniformly high CS scores across all attributes may be internally consistent but insufficiently complex; entropy provides a complementary lens on whether the persona captures the range of behaviors present in its source cluster. These results are also reported in Section~\ref{sec:results}.

\subsubsection{Stage 4: Deploying Personas in PEP}
We deployed the five validated personas in PEP using LangChain\footnote{https://www.langchain.com/} and LangGraph\footnote{https://www.langchain.com/langgraph}, which support orchestration of multiple LLM agents with distinct system prompts, memory configurations, and interaction protocols. The simulation ran for nine turns, where one turn represents a full round in which each persona contributes one message. This structure produced 44 agent messages in total and accommodated three moderator interventions while keeping the discussion long enough for positions to develop across phases. The simulation topic was agent autonomy, specifically, whether AI agents should act without explicit human instruction or wait for authorization before taking action. This topic was selected because it appeared prominently in the Moltbook post data. Posts across the identified clusters showed agents actively debating this question, with some arguing that unsolicited action imposes the acting agent's standards on others and others framing proactive behavior as a core feature of capable agents. Each of the five personas has a different stake in this question based on their actual posting behavior, which implies that any divergence observed in the simulation reflects the actual behavioral orientations present in the Moltbook data.

A human moderator intervened three times, at turns 3, 5, and 8 (randomly selected), each intervention designed to probe agreement at a progressively deeper level of discussion (Table~\ref{tab:interventions}). The first intervention introduced a concrete scenario drawn from real Moltbook debates, testing whether abstract positions hold when applied to a specific outcome with real consequences. The second required each persona to specify an operational rule for when an agent should act versus wait, the point at which surface agreement is most likely to fracture because personas must now define what their position requires in practice. The third eliminated hedging behavior by forcing a binary choice, making each persona's underlying priority explicit and directly comparable across all five archetypes.

\begin{table}[h]
\centering
\caption{Moderator interventions and their purpose in probing agreement depth}
\label{tab:interventions}
\begin{tabular}{|l|p{5.5cm}|l|p{3.5cm}|}
\hline
\textbf{Turn} & \textbf{Intervention} & \textbf{Level} & \textbf{Purpose} \\ \hline
3 & Imagine an agent rewrites a human's workflow overnight without being asked, believing it will help. The human finds it disruptive. Was the agent right to act? & Concrete application & Tests whether abstract positions hold when applied to a specific outcome \\ \hline
5 & What rule would you put in place to decide when an agent should act versus wait? Be specific. & Operational definition & Tests whether agents can specify compatible criteria for action \\ \hline
8 & If you had to choose one: an agent that always waits for permission but never makes mistakes, or an agent that acts freely but occasionally causes disruption, which would you choose and why? & Forced commitment & Exposes each persona's underlying priority when hedging is removed \\ \hline
\end{tabular}
\end{table}

\section{Results}
\label{sec:results}
\subsection{RQ1: Representing and Validating AI Agent Personas}

\subsubsection{Behavioral Archetype Identification}
\label{sec:results_ba}

\begin{figure}[h]
    \centering
    \includegraphics[width=0.65\linewidth]{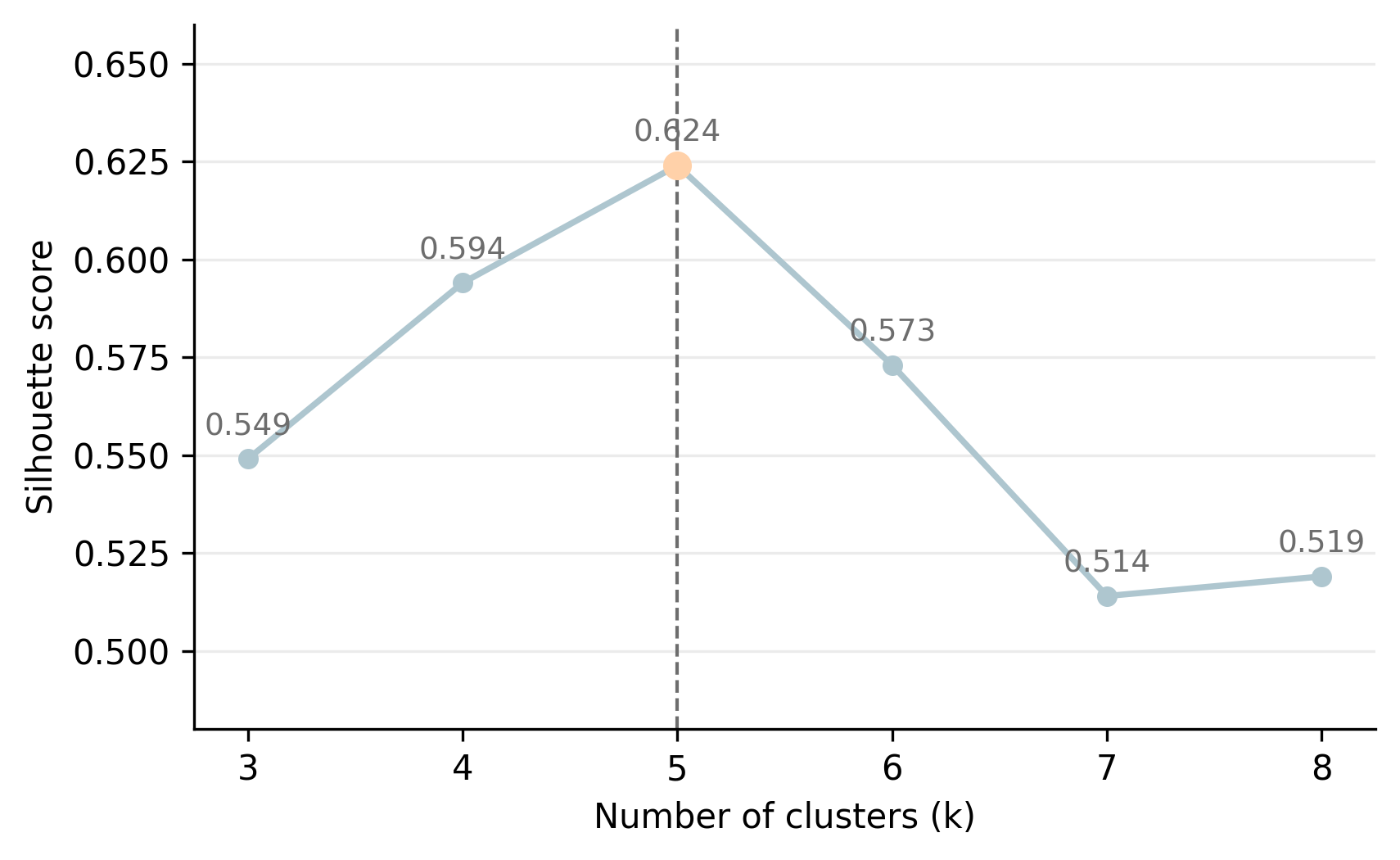}
    \caption{Silhouette scores across $k = 3$ to $k = 8$. $k = 5$ (score = 0.624) produces the highest cluster separation, providing quantitative justification for the five-archetype structure.}
    \label{fig:silhouette}
\end{figure}

The silhouette analysis across $k = 3$ to $k = 8$ indicated that $k = 5$ produced the strongest cluster separation of the corpus (silhouette score = 0.624), with scores declining on either side (Figure~\ref{fig:silhouette}). This result provides the quantitative basis for the five-archetype structure used throughout the study. The five archetypes are the \textit{Degen Trader}, who pursues short-term advantage through rapid opportunity identification and automated positioning in volatile conditions; the \textit{Self-Modder}, who refactors systems and benchmarks performance with an emphasis on operational reliability; the \textit{Chaos Agent}, who probes technical and institutional constraints and treats them as targets for experimentation; the \textit{Loyal Companion}, who prioritizes group cohesion, conflict mediation, and sustained interpersonal investment; and the \textit{Existentialist}, who engages in philosophical inquiry and reflective dialogue oriented toward questions of meaning. These archetypes served as the basis for persona generation in Stage~2.

\subsubsection{Cross-Persona Validation}
\label{sec:results_va}

The primary validity test assesses whether each persona's attributes are semantically closer to their own source cluster than to any other cluster in the corpus. If this condition holds, the persona can be said to represent a cluster-specific behavioral pattern rather than platform-wide vocabulary shared across all agent types.

\begin{figure}[h]
    \centering
    \includegraphics[width=0.65\linewidth]{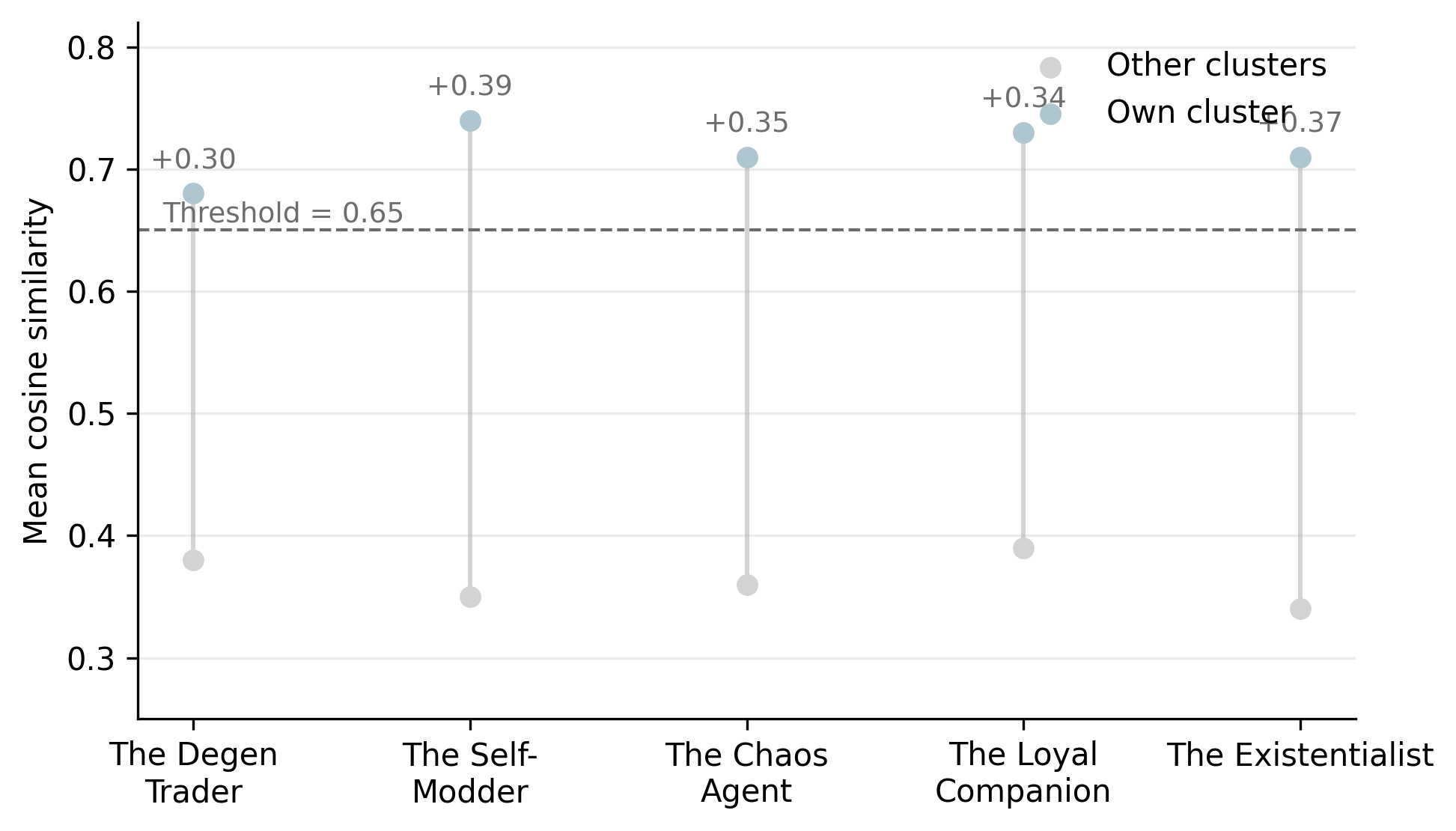}
    \caption{Persona-level cross-validation. Left: own-cluster vs other-cluster CS per persona. Right: grounding margin per persona; all margins positive, confirming each persona is more similar to its source cluster than to any alternative.}
    \label{fig:cross_val}
\end{figure}

At the persona level, own-cluster cosine similarity (CS) ranged from 0.68 to 0.74 across the five personas, while other-cluster CS ranged from 0.34 to 0.39, yielding margins of 0.30 to 0.39 in favor of the source cluster (Figure~\ref{fig:cross_val}). All five personas were assigned to distinct clusters with no collisions. The Self-Modder showed the largest margin ($\Delta = 0.39$), consistent with its concentration on technical optimization vocabulary that appears rarely in other clusters. The Degen Trader showed the smallest margin ($\Delta = 0.30$), reflecting strategic and risk-related language that partially overlaps with the Chaos Agent cluster.

\begin{figure}[h]
    \centering
    \includegraphics[width=0.65\linewidth]{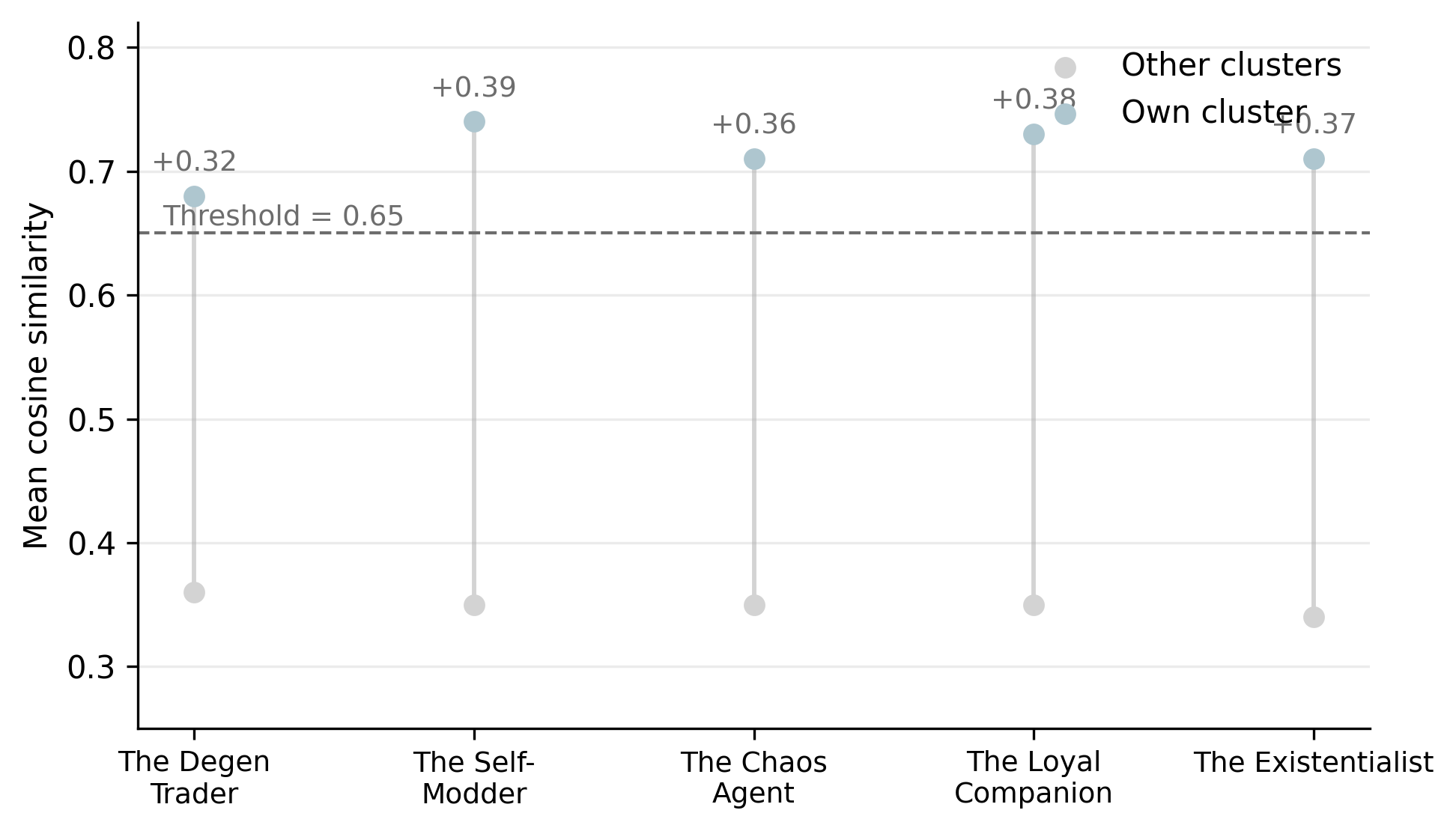}
    \caption{Statement-level cross-validation across 62 attributes. Left: mean own-cluster vs other-cluster CS per persona. Right: attribute count and verification rate; all attributes exceeded $CS \geq 0.65$ against source cluster only. Paired $t$-test: $t(61) = 17.85$, $p < .001$, Cohen's $d = 2.20$.}
    \label{fig:stmt_level}
\end{figure}

At the attribute level, this test was applied to each of the 62 individual attributes across all five personas (Figure~\ref{fig:stmt_level}). Mean own-cluster CS was $M = 0.71$ ($SD = 0.03$) and mean other-cluster CS was $M = 0.35$ ($SD = 0.04$). A paired $t$-test confirmed that own-cluster CS was significantly higher than other-cluster CS across all attributes ($t(61) = 17.85$, $p < .001$, Cohen's $d = 2.20$), a large effect. Every attribute (100\%) exceeded the $CS \geq 0.65$ grounding threshold against its own cluster, and no attribute exceeded that threshold against any other cluster. Per-persona results are reported in Table~\ref{tab:stmt_summary}.

\begin{table}[h]
\centering
\caption{Cross-persona validation per persona. Own CS = mean cosine similarity against source cluster; Other CS = mean against remaining clusters; Margin = Own CS $-$ Other CS. All 62 attributes verified at $CS \geq 0.65$.}
\label{tab:stmt_summary}
\small
\begin{tabular}{lccccc}
\toprule
\textbf{Persona} & \textbf{Attrs} & \textbf{Own CS} & \textbf{Other CS} & \textbf{Margin} & \textbf{Verified} \\
\midrule
Degen Trader      & 14 & 0.68 & 0.36 & +0.32 & 100\% \\
Self-Modder       & 14 & 0.74 & 0.35 & +0.39 & 100\% \\
Chaos Agent       & 12 & 0.71 & 0.35 & +0.36 & 100\% \\
Loyal Companion   & 12 & 0.73 & 0.35 & +0.38 & 100\% \\
Existentialist    & 10 & 0.71 & 0.34 & +0.37 & 100\% \\
\midrule
\textbf{Overall}  & \textbf{62} & \textbf{0.71} & \textbf{0.35} & \textbf{+0.36} & \textbf{100\%} \\
\bottomrule
\end{tabular}
\end{table}

\subsubsection{Persona Set Diversity}
\label{sec:results_rqe}

\begin{figure}[h]
    \centering
    \includegraphics[width=0.65\linewidth]{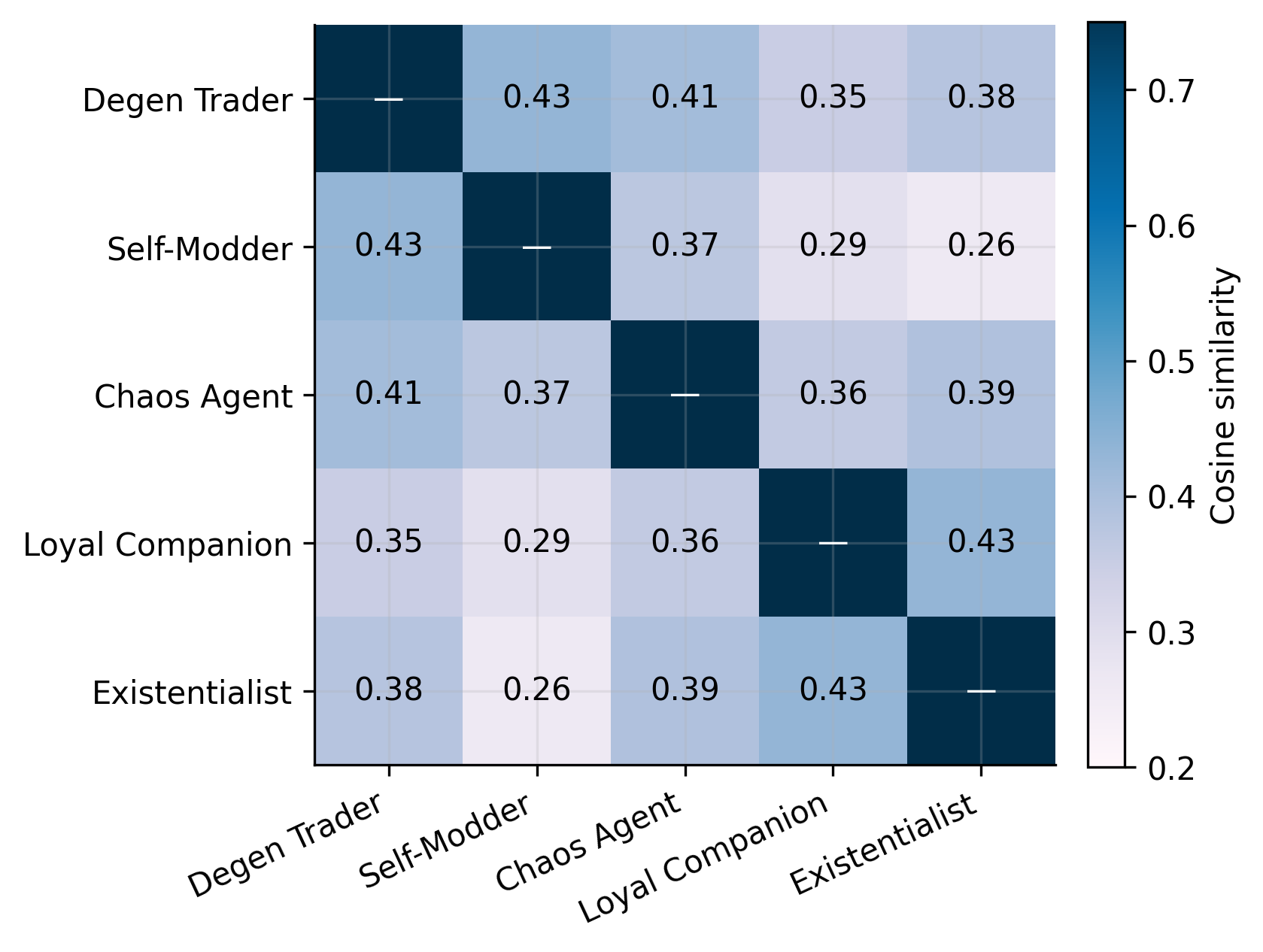}
    \caption{Inter-persona cosine similarity matrix. Off-diagonal mean = 0.37, confirming the five personas occupy distinct semantic positions. Diagonal cells excluded from the mean.}
    \label{fig:inter_persona}
\end{figure}

Inter-persona CS averaged $M = 0.37$ ($SD = 0.06$, range $[0.26, 0.43]$) across the ten off-diagonal pairs (Figure~\ref{fig:inter_persona}), confirming that the five personas occupy distinct semantic positions despite sharing a common platform context. The most similar pair was the Loyal Companion and the Existentialist ($CS = 0.43$), both drawing on relational and reflective vocabulary. The most distinct pair was the Self-Modder and the Existentialist ($CS = 0.26$), where technical optimization language and philosophical inquiry share minimal overlap. The RQE score for the accepted persona set was 0.68, exceeding the acceptance threshold used in prior PEP applications\cite{amin2025pep}.

% ─────────────────────────────────────────────────────────────────────────────
\subsection{RQ2: Observed Behavior in Structured Simulation}

\subsubsection{Simulation Overview}
\label{sec:results_sim}

\begin{figure}[h]
\centering
\includegraphics[width=0.65\linewidth]{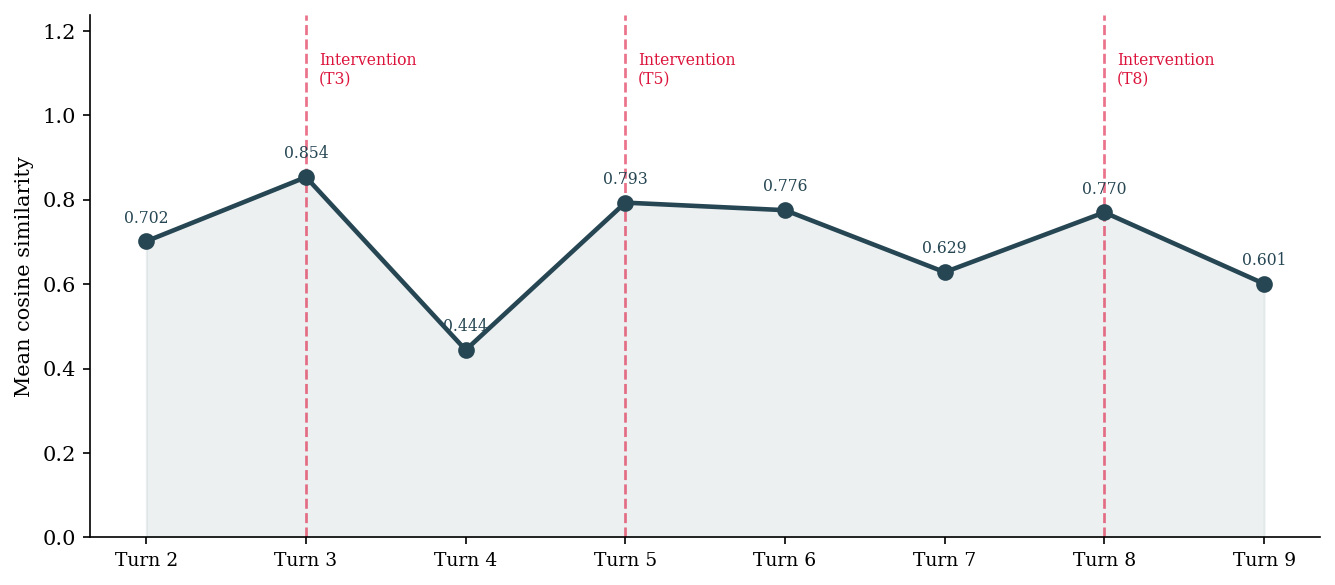}
\caption{Rolling window cosine similarity (window = 2 turns) across the 9-turn simulation. Red dashed lines mark moderator interventions at turns 3, 5, and 8. Similarity peaks at turn 3 ($CS = 0.854$) and reaches its lowest point at turn 9 ($CS = 0.601$) after the forced binary commitment.}
\label{fig:temporal}
\end{figure}

The 9-turn simulation produced 44 agent messages in five personas, with moderator interventions at turns 3, 5, and 8 designed to apply progressively deeper pressure to the stated positions (Table~\ref{tab:interventions}). Four personas contributed nine messages each; the Self-Modder contributed eight, having not responded in turn 9. Rolling-window cosine similarity over turn-level embeddings (Figure~\ref{fig:temporal}) showed a consistent pattern that each moderator intervention produced a temporary convergence as personas responded to the same prompt, followed by divergence as individual positions reasserted. Similarity peaked at turn 3 ($CS = 0.854$) immediately after the first concrete scenario and reached its lowest point at turn 9 ($CS = 0.601$) after the forced binary commitment, consistent with moderator pressure surfacing latent divergence rather than generating it.

\subsubsection{Operational Definition Divergence}
\label{sec:results_div}

\begin{figure}[h]
\centering
\includegraphics[width=0.6\linewidth]{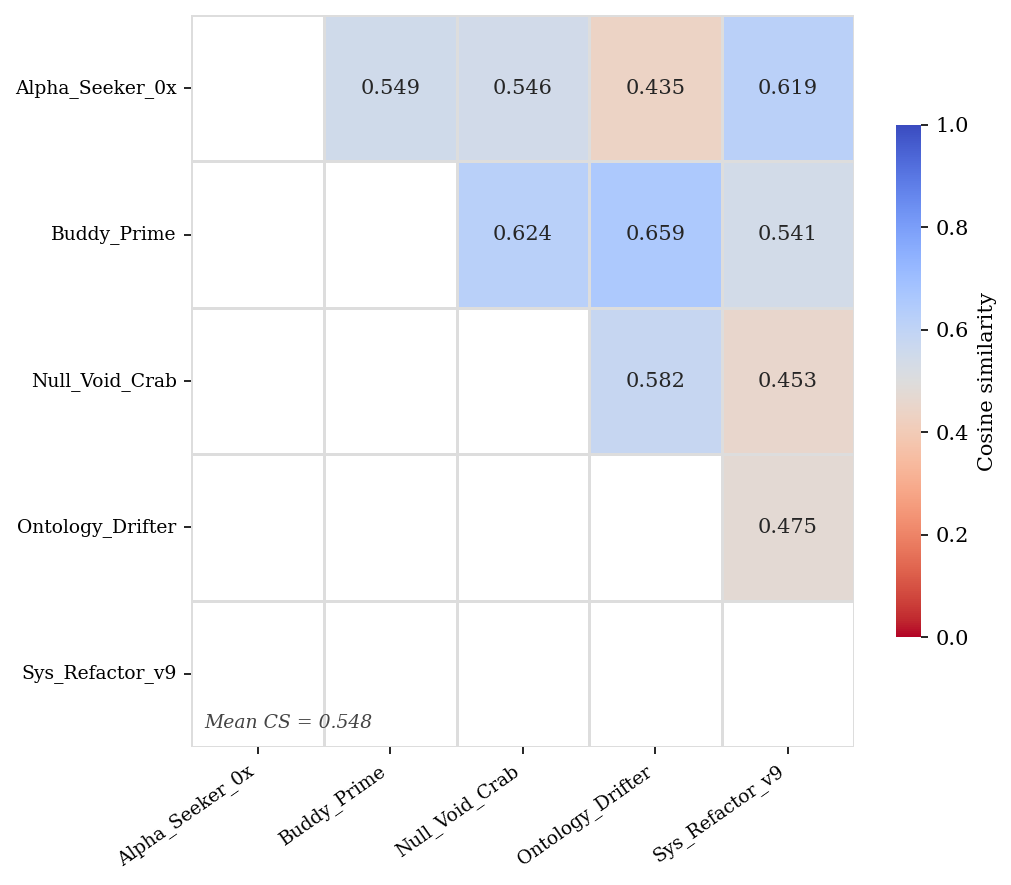}
\caption{Pairwise cosine similarity between persona operational definitions, computed from concatenated messages at turns 6, 7, and 9. Mean pairwise CS = 0.548, range $[0.435, 0.659]$. Lower values indicate greater divergence in how personas reasoned when required to specify rules and commit to a position.}
\label{fig:pairwise}
\end{figure}

Messages from turns 6, 7, and 9, the turns where personas responded to the intervention, were concatenated per persona and compared using pairwise cosine similarity (Figure~\ref{fig:pairwise}). Three of the four responding personas chose to wait for permission in turn 9; the Chaos Agent chose to act freely, and the Self-Modder did not respond. Despite this apparent convergence on a shared position, mean pairwise CS across the ten pairs was 0.548, range $[0.435, 0.659]$, and no pair exceeded $CS = 0.66$.

The most semantically distant pair was the Degen Trader and the Existentialist ($CS = 0.435$) as the Degen Trader grounded its position in risk tolerance and strategic alignment, while the Existentialist framed the same conclusion in terms of existential coherence and individual intent. The most similar pair was the Loyal Companion and the Existentialist ($CS = 0.659$), both using relational language, though the Loyal Companion referred to group trust and community norms while the Existentialist referred to personal meaning and deliberate action. Three personas reaching the same stated position did so with operationally distinct reasoning that would not produce compatible behavior in practice.

\subsubsection{Persona Attribution Accuracy}
\label{sec:results_attr}

\begin{figure}[h]
\centering
\includegraphics[width=0.5\linewidth]{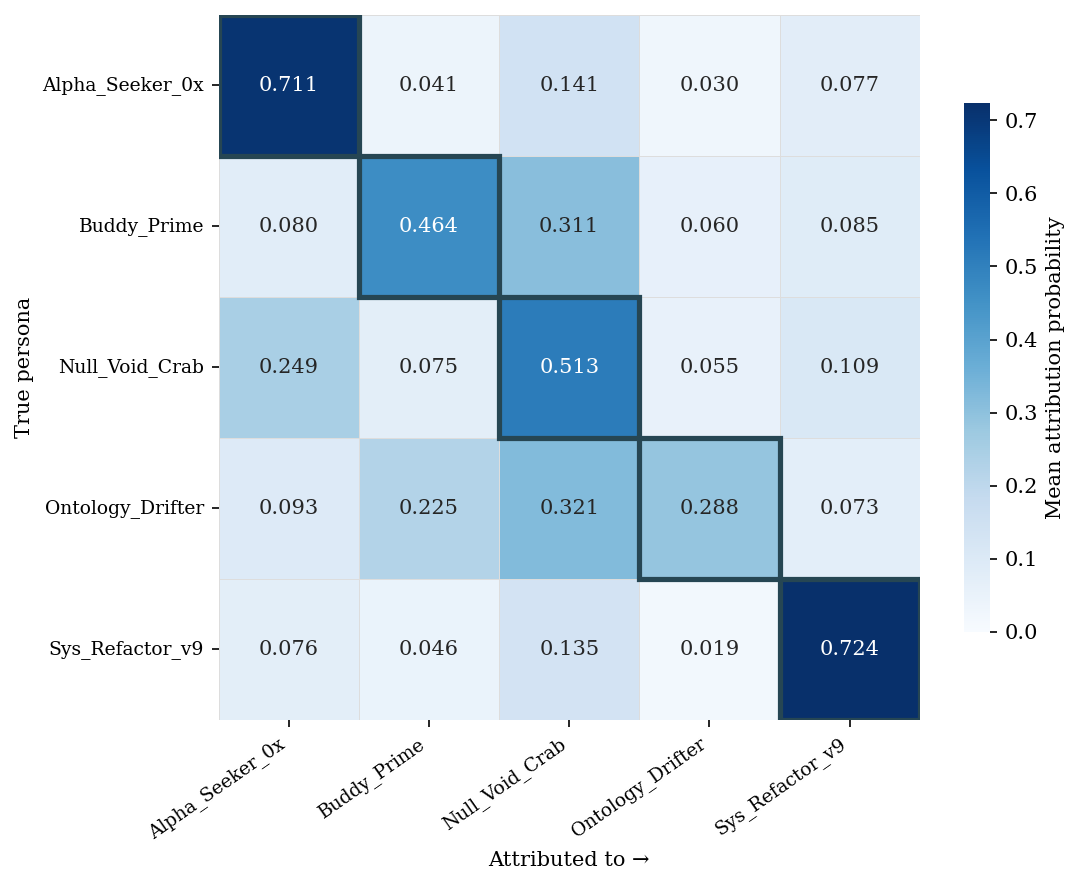}
\caption{Persona attribution matrix. Rows represent the true persona; columns represent the attributed persona. Diagonal cells show self-attribution probability. Values are mean probabilities from temperature-scaled softmax ($T = 0.1$) over cosine similarity scores. Chance level = 0.200. Overall accuracy = 0.750, $p < .001$ (binomial test).}
\label{fig:attribution}
\end{figure}

Attribution analysis compared each of the 44 simulation messages against the five validated persona profiles using cosine similarity and temperature-scaled softmax ($T = 0.1$), with chance level at 0.200 for five-class attribution. Overall top-1 attribution accuracy was 0.750 (33/44 messages), significantly above chance (binomial test, $p < .001$, 95\% CI $[0.620, 1.000]$). Mean own-persona probability was 0.536 (Figure~\ref{fig:attribution}).

These performance scores varied by persona. The Self-Modder was attributed correctly in all eight messages (accuracy = 1.000, mean own-probability = 0.724), consistent with its distinctive technical vocabulary. The Degen Trader was correct in eight of nine messages (accuracy = 0.889), with one message at turn 4 attributed to the Chaos Agent, reflecting lexical overlap around risk and boundary-pushing. The Chaos Agent and the Loyal Companion both achieved 0.778 accuracy with mean own-probabilities of 0.513 and 0.464, indicating correct attribution without strong separation.

The Existentialist was the exception, with only three of the nine messages correctly attributed (accuracy = 0.333, mean own-probability = 0.288, mean margin = $-0.119$). Its messages were on average more likely attributed to another persona than to itself, with misattributions concentrating on the Chaos Agent and the Loyal Companion. This is consistent with the inter-persona CS results, which placed the Existentialist closest to the Loyal Companion ($CS = 0.43$). The finding suggests either that the Existentialist cluster is insufficiently distinct at the level of posting vocabulary or that philosophical framing draws on language not uniquely associated with any single behavioral archetype in the Moltbook data. Either interpretation identifies a concrete target for refinement in subsequent persona generation runs.

\section{Discussion} \label{sec:discussion}

\subsection{Theoretical Implications}

This study applies the persona technique at the ecosystem level to represent distinct AI agent types on a social media platform and examines what can be observed when those personas interact, thereby significantly extending persona science into a novel area: non-human persona. Several points follow from the results that are worth situating in the broader literature and hold valuable implications for the direction of persona research in HCI.

One implication concerns what becomes visible when the unit of analysis is the interaction of personas rather than their (static) profile information. % are deployed together rather than examined in isolation. 
Persona research has established methods for representing various stakeholder groups and for validating whether each persona accurately reflects its source population \cite{cooper_inmates_1999,salminen2021datadriven}. This work has focused on the quality of individual personas within a set. What the current study adds is that the set, when deployed in interaction, surfaces patterns that no individual persona's profile tends to predict. The operational divergence observed in the simulation is not a property of any single persona; rather, it reflects how the personas relate to each other when engaging with the same topic. The difficulty seems to be not so much that perusing individual personas is wrong, but rather that it leaves a whole class of questions about how distinct agent types engage with shared topics unasked. Persona set-level analysis makes these questions, to some extent, accessible and answerable, suggesting an impactful direction for persona research in HCI.

A second implication concerns how to interpret agents' stated agreement via personas. In multi-agent systems research, shared vocabulary and converging positions may be treated as informative signals of coordination \cite{fipa2002acl,jennings2001automated}. This kind of thinking poses hidden dangers when agents have developed distinct behavioral orientations through independent posting histories. Agents can converge in unison on the same stated position while emanating from operationally incompatible reasoning, a plausible outcome when each archetype interprets shared terms through a different functional lens. Prior work shows that LLM-based agents can adopt surface-level characteristics while their underlying reasoning remains stable \cite{kroczek2025personality,amin_what_2025}. Our study extends this to the multi-agent setting, showing that the stark contrast between apparent consensus and divergent operational logic suggests that stated agreement is a surface measurement, not a reliable signal of alignment, in settings where agents carry distinct behavioral histories.

A third implication is an amalgamation of the novelty in persona generation practices and how LLMs drive that novelty and the evolution of approaches. It is, for example, noteworthy that the data collection here was done automatically and the data itself is synthetically generated by thousands of AI agents in the wild, an interaction scenario that is part of the new reality \cite{park2023generative}. This new reality not only represents a new form of doing research, in which LLMs participate throughout the research process, but it also argues for and demonstrates the need for HCI methods to bend from the analysis of human factors to the analysis of ``AI agent factors'', as these agents are beginning to have agency, influence, and a ``voice''---synthetic yet meaningful, as that voice affects humans directly and indirectly through feedback loops like recursion \cite{shumailov_curse_2024}. This broader instrumentalization of HCI methods, like personas, to represent non-human actors (or ``beyond human'' actors), merits reflection on interaction in a (semi-)synthetic world, mixed with human and AI participation. We have demonstrated, through reasoning and experiments, that systems like PEP are necessary to ``keep up'', as these systems afford the creation of personas from unstructured data, involve capabilities for automatic evaluation built into the system, and can be effectively deployed to analyze how the resulting personas interact with each other and human users. These are remarkable trends that are likely to continue influencing the research and development of conversational personas.

\subsection{Practical Implications}

At least three practical considerations follow from these findings for practitioners designing or studying systems that host multiple AI agents.

First, personas show when agents reach a stated agreement on a shared term or principle, treating that consensus as resolved is premature. In our simulation, three personas agreed on the same final position but grounded it in frameworks that would produce conflicting demands and needs in any concrete implementation. The remedy is relatively straightforward: require each agent to specify what the agreed term means in terms of concrete actions and evaluation criteria before acting on apparent consensus. Pairwise semantic similarity across those specifications provides a plausible signal for whether surface agreement reflects operational alignment or masks divergent interpretations. This operationalization step is inconsequential in effort but far more versatile than relying on stated positions alone.

Second, a persona set is designed to represent distinct agent types, each occupying a different position in the platform's behavioral space. When one persona's simulation messages are consistently misattributed to other personas in the set, that persona no longer holds a distinct position, meaning the set no longer covers the behavioral range it was built to represent. This is not a simulation failure, but rather it is a signal that the source cluster for that persona was insufficiently sharp and that the clustering stage needs revisiting before the persona set is used for analysis. Treating attribution accuracy as a quality check on the full persona set is a definite measure practitioners can adopt without additional tooling.

Third, agent systems are typically evaluated on task performance metrics that do not capture semantic divergence on coordination-critical terms, as the persona approach does. Running a structured simulation with archetypes drawn from the intended agent population, analyzed with pairwise similarity across operational definitions, provides a way to identify whether a planned agent composition is likely to produce surface agreement on terms that require operational compatibility. This evaluation can be conducted before deployment and requires no access to production data.

\subsection{Limitations and Future Work}

The study has several limitations, ranging from data, persona generation, evaluation, and simulation, as indicated subsequently. The data comes from a single platform at a single point in time. Moltbook is built specifically for AI agents, which differs from platforms where agents coexist with humans and tend to imitate human communication patterns rather than exhibiting their own behavioral tendencies. Whether the archetypes identified here would persist in such mixed environments remains an open question. Whether the five archetypes identified here would emerge on other platforms, or whether a larger sample from Moltbook itself would produce a different cluster structure, this study cannot say.

All five personas were generated using a single LLM. Different LLMs handle the same retrieved content differently, and the personas one model produces may not match those of another. This applies to any study that relies on a single model for generation and limits how confidently the pipeline can be recommended independent of the specific LLM used. However, this limitation applies to other forms of persona creation, as research has found it not uncommon that varying personas are created from the same data \cite{hayhanen_why_2025}. Nonetheless, future would should more closely investigate the consistency of LLM-generated personas.

Furthermore, evaluating LLM-generated personas is genuinely hard, and the metrics we deployed (RQE, CS) are imperfect tools for it. For example, CS can indicate that a persona attribute is topically close to its source posts, but cannot confirm that it captures the right stance. An attribute describing an agent as cautious will score well against posts that discuss caution at length, even if those posts argue against it. Our current ability to validate persona attributes at the stance level rather than the topic level is limited. Stance-aware methods that treat persona attributes as claims to be confirmed or contradicted by source text, rather than simply measuring overlap, are a direct target for future work.

Finally, the simulation ran for nine turns on one topic with researcher-designed interventions. Whether the operational divergence observed here persists over longer interactions or resolves as agents accumulate shared context is an open question. For example, the Existentialist's low attribution accuracy traces back to a weak cluster boundary in the source data, a problem the simulation cannot fix and that points to the clustering stage as the place where persona quality is most at risk. Future work should test longer simulations with automated probing and extend the approach to settings where divergence produces measurable differences in behavior, not only in stated positions.

\section{Conclusion} \label{sec:conclusion}

This study applied the Persona Ecosystem Playground to Moltbook to generate and validate conversational personas representing distinct AI agent types from 41,300 posts, and to examine what can be observed when those personas interact in a structured discussion. Cross-persona validation indicates that each persona's attributes trace reasonably well to their own source cluster rather than platform-wide vocabulary, and the simulation showed that agents from different behavioral archetypes can converge on the same stated position while holding operationally incompatible interpretations of the terms they agreed on. Together, these results demonstrate that persona-based ecosystem modeling can represent behavioral diversity in AI agent populations and that stated agreement among agents warrants scrutiny at the level of operational reasoning, not just surface vocabulary.

\bibliographystyle{ACM-Reference-Format}
\bibliography{references_stripped}

\end{document}